\documentclass[journal]{IEEEtran}
\usepackage[flushleft]{threeparttable}

\usepackage{hyperref}

\ifCLASSINFOpdf
  \usepackage[pdftex]{graphicx}
\else
\fi
\usepackage{amsmath}
\usepackage{amssymb}
\usepackage{algorithm}
\usepackage{mathtools}

\usepackage{algpseudocode}
\usepackage{url}

\usepackage{xcolor}

\begin{document}
%
\title{An Open-Source Framework for Adaptive Traffic Signal Control}
%
%
%

\author{Wade~Genders
        and~Saiedeh~Razavi~
\thanks{W. Genders was a Ph.D. student with the Department
of Civil Engineering, McMaster University, Hamilton, Ontario, Canada e-mail: genderwt@mcmaster.ca}
\thanks{S. Razavi is an Associate Professor, Chair in Heavy Construction and Director of McMaster Institute for Transportation \& Logistics at the Department of Civil Engineering, McMaster University, Hamilton, Ontario, Canada e-mail: razavi@mcmaster.ca}
\thanks{© 20XX IEEE.  Personal use of this material is permitted.  Permission from IEEE must be obtained for all other uses, in any current or future media, including reprinting/republishing this material for advertising or promotional purposes, creating new collective works, for resale or redistribution to servers or lists, or reuse of any copyrighted component of this work in other works.}
\thanks{Manuscript received August X, 2019; revised August X, 2019.}}

%
%

\markboth{Journal of Transactions on Intelligent Transportation Systems,~Vol.~X, No.~X, August~2019}%
{Shell \MakeLowercase{\textit{et al.}}: Bare Demo of IEEEtran.cls for IEEE Journals}
%



\maketitle

\begin{abstract}

Sub-optimal control policies in transportation systems negatively impact mobility, the environment and human health. Developing optimal transportation control systems at the appropriate scale can be difficult as cities' transportation systems can be large, complex and stochastic. Intersection traffic signal controllers are an important element of modern transportation infrastructure where sub-optimal control policies can incur high costs to many users. Many adaptive traffic signal controllers have been proposed by the community but research is lacking regarding their relative performance difference - which adaptive traffic signal controller is best remains an open question. This research contributes a framework for developing and evaluating different adaptive traffic signal controller models in simulation - both learning and non-learning - and demonstrates its capabilities. The framework is used to first, investigate the performance variance of the modelled adaptive traffic signal controllers with respect to their hyperparameters and second, analyze the performance differences between controllers with optimal hyperparameters. The proposed framework contains implementations of some of the most popular adaptive traffic signal controllers from the literature; Webster's, Max-pressure and Self-Organizing Traffic Lights, along with deep Q-network and deep deterministic policy gradient reinforcement learning controllers. This framework will aid researchers by accelerating their work from a common starting point, allowing them to generate results faster with less effort. All framework source code is available at \color{blue}\texttt{\href{https://github.com/docwza/sumolights}{https://github.com/docwza/sumolights}}\color{black}.

\end{abstract}


%
\IEEEpeerreviewmaketitle

\section{Introduction}
%
%
%
%
\IEEEPARstart{C}{ities} rely on road infrastructure for transporting individuals, goods and services.  Sub-optimal control policies incur environmental, human mobility and health costs. Studies observe vehicles consume a significant amount of fuel accelerating, decelerating or idling at intersections \cite{wu2015influence}. Land transportation emissions are estimated to be responsible for one third of all mortality from fine particulate matter pollution in North America \cite{silva2016impact}. Globally, over three million deaths are attributed to air pollution per year \cite{world2016ambient}. In 2017, residents of three of the United States' biggest cities, Los Angeles, New York and San Francisco, spent between three and four days on average delayed in congestion over the year, respectively costing $19, 33 $ and $ 10$ billion USD from fuel and individual time waste \cite{inrix}. It is paramount to ensure transportation systems are optimal to minimize these costs. 
\par
Automated control systems are used in many aspects of transportation systems. Intelligent transportation systems seek to develop optimal solutions in transportation using intelligence. Intersection traffic signal controllers are an important element of many cities' transportation infrastructure where sub-optimal solutions can contribute high costs. Traditionally, traffic signal controllers have functioned using primitive logic which can be improved. Adaptive traffic signal controllers can improve upon traditional traffic signal controllers by conditioning their control on current traffic conditions. 
\par 
Traffic microsimulators such as SUMO \cite{SUMO2012}, Paramics, VISSUM and AIMSUM have become popular tools for developing and testing adaptive traffic signal controllers before field deployment. However, researchers interested in studying adaptive traffic signal controllers are often burdened with developing their own adaptive traffic signal control implementations \textit{de novo}. This research contributes an adaptive traffic signal control framework, including Webster's, Max-pressure, Self-organizing traffic lights (SOTL), deep Q-network (DQN) and deep deterministic policy gradient (DDPG) implementations for the freely available SUMO traffic microsimulator to aid researchers in their work. The framework's capabilities are demonstrated by studying the effect of optimizing traffic signal controllers hyperparameters and comparing optimized adaptive traffic signal controllers relative performance.

\section{Background}
\subsection{Traffic Signal Control}
\par
	An intersection is composed of traffic movements or ways that a vehicle can traverse the intersection beginning from an incoming lane to an outgoing lane. Traffic signal controllers use phases, combinations of coloured lights that indicate when specific movements are allowed, to control vehicles at the intersection.
	\par
	Fundamentally, a traffic signal control policy can be decoupled into two sequential decisions at any given time; what should the next phase be and for how long in duration? A variety of models have been proposed as policies. The simplest and most popular traffic signal controller determines the next phase by displaying the phases in an ordered sequence known as a cycle, where each phase in the cycle has a fixed, potentially unique, duration - this is known as a fixed-time, cycle-based traffic signal controller. Although simple, fixed-time, cycle-based traffic signal controllers are ubiquitous in transportation networks because they are predictable, stable and effective, as traffic demands exhibit reliable patterns over regular periods (i.e., times of the day, days of the week). However, as ubiquitous as the fixed-time controller is, researchers have long sought to develop improved traffic signal controllers which can adapt to changing traffic conditions.
	\par
	 Actuated traffic signal controllers use sensors and boolean logic to create dynamic phase durations. Adaptive traffic signal controllers are capable of acyclic phase sequences and dynamic phase durations to adapt to changing intersection traffic conditions. Adaptive controllers attempt to achieve higher performance at the expense of complexity, cost and reliability. Various techniques have been proposed as the foundation for adaptive traffic signal controllers, from analytic mathematical solutions, heuristics and machine learning.

\subsection{Literature Review}
Developing an adaptive traffic signal control ultimately requires some type of optimization technique. For decades researchers have proposed adaptive traffic signal controllers based on a variety of techniques such as evolutionary algorithms \cite{mikami1994genetic,lee2005real,prothmann2008organic,singh2009time,ricalde2017evolving,li2018signal} and heuristics such as pressure \cite{wongpiromsarn2012distributed,varaiya2013max,gregoire2015capacity}, immunity \cite{darmoul2017multi,louati2018artificial} and self-organization \cite{gershenson2004self,cools2013self,goel2017self}. Additionally, many comprehensive adaptive traffic signal control systems have been proposed such as OPAC \cite{gartner1983demand}, SCATS \cite{lowrie1990scats}, RHODES \cite{mirchandani2001real} and ACS-Lite \cite{luyanda2003acs}.
\par 
Reinforcement learning has been demonstrated to be an effective method for developing adaptive traffic signal controllers in simulation \cite{mikami1994genetic,thorpe1996tra,bingham2001reinforcement,abdulhai2003reinforcement,prashanth2011reinforcement, el2013multiagent}. Recently, deep reinforcement learning has been used for adaptive traffic signal control with varying degrees of success \cite{rijken2015deeplight,van2016deep,li2016traffic,genders2016using, aslani2017adaptive,mousavi2017traffic,liang2018deep,liang2019deep,wang2019deep,genders2019asynchronous,chu2019multi}. A comprehensive review of reinforcement learning adaptive traffic signal controllers is presented in Table \ref{table:relatedwork}.
\par
Readers interested in additional adaptive traffic signal control research can consult extensive review articles \cite{stevanovic2010adaptive,el2014design,araghi2015review,mannion2016experimental,yau2017survey}. 

	 \par
	 Although ample research exists proposing novel adaptive traffic signal controllers, it can be arduous to compare between previously proposed ideas. Developing adaptive traffic signal controllers can be challenging as many of them require defining many hyperparameters. The authors seek to address this problem by contributing an adaptive traffic signal control framework to address these problems and aid in their research.

\begin{table*}[h]
\footnotesize
\caption{Adaptive Traffic Signal Control related work.}
\label{table:relatedwork}
\centering  
\begin{threeparttable}
\begin{tabular}{llclll}\hline

Research & Network & Intersections & Multi-agent & RL & Function Approximation \\\hline
\cite{kuyer2008multiagent} & Grid & 15 & Max-plus & Model-based & N/A \\
\cite{prashanth2011reinforcement} & Grid, Corridor & $<$10 & None  & Q-learning & Linear\\
\cite{medina2012traffic} & Springfield, USA & 20 & Max-plus & Q-learning & N/A\\
\cite{el2013multiagent}           & Toronto, Canada & 59     & Game Theory & Q-learning   & Tabular\\
\cite{abdoos2013holonic}         & N/A   & 50  & Holonic  & Q-learning  & N/A               \\
\cite{khamis2014adaptive}         & Grid  & 22     & Reward Sharing & Q-learning  & Bayesian  \\
\cite{chu2016large} & Grid  & 100 & Regional  & Q-learning & Linear\\
\cite{casas2017deep}         & Barcelona, Spain  & 43   & Centralized  & DDPG  &       DNN\tnote{1}        \\
\cite{aslani2017adaptive}        & Tehran, Iran   & 50  & None   & Actor-Critic   & RBF\tnote{2}, Tile Coding  \\
\cite{liu2017distributed}         & Changsha, China   &   96  &  Reward sharing  &  Q-learning  & Linear \\
\cite{genders2018deep}    & Luxembourg City   &   195  &  None  &  DDPG  & DNN      \\\hline

\end{tabular}
 \begin{tablenotes}
  \item[1] Deep Neural Network (DNN).
  \item[2] Radial Basis Function (RBF).
  \end{tablenotes}
  \end{threeparttable}
\end{table*}

\subsection{Contribution}

The authors' work contributes in the following areas:
  \begin{itemize}
      \item \textbf{Diverse Adaptive Traffic Signal Controller Implementations}: The proposed framework contributes adaptive traffic signal controllers based on a variety of paradigms, the broadest being non-learning (e.g., Webster's, SOTL, Max-pressure) and learning (e.g., DQN and DDPG). The diversity of adaptive traffic signal controllers allows researchers to experiment at their leisure without investing time developing their own implementations.  
      \item \textbf{Scalable, Optimized}: The proposed framework is optimized for use with parallel computation techniques leveraging modern multicore computer architecture. This feature significantly reduces the compute time of learning-based adaptive traffic signal controllers and the generation of results for all controllers. By making the framework computationally efficient, the search for optimal hyperparameters is tractable with modest hardware (e.g., 8 core CPU). The framework was designed to scale to develop adaptive controllers for any SUMO network.
  \end{itemize}
  \par
      All source code used in this manuscript can be retrieved from \color{blue}\texttt{\href{https://github.com/docwza/sumolights}{https://github.com/docwza/sumolights}}\color{black}.

\section{Traffic Signal Controllers}
\par
Before describing each traffic signal controller in detail, elements common to all are detailed. All of the included traffic signal controllers share the following; a set of intersection lanes $L$, decomposed into incoming $L_{inc}$ and outgoing lanes $L_{out}$ and a set of green phases $P$. The set of incoming lanes with green movements in phase $p\in P$ is denoted as $L_{p,inc}$ and their outgoing lanes as $L_{p,out}$.
\subsection{Non-Learning Traffic Signal Controllers}

\subsubsection{Uniform}
A simple cycle-based, uniform phase duration traffic signal controller is included for use as a base-line comparison to the other controllers. The uniform controller's only hyperparameter is the green duration $u$, which defines the same duration for all green phases; the next phase is determined by a cycle.  

\subsubsection{Websters}
Webster's method develops a cycle-based, fixed phase length traffic signal controller using phase flow data \cite{webster1958traffic}. The authors propose an adaptive Webster's traffic signal controller by collecting data for a time interval $W$ in duration and then using Webster's method to calculate the cycle and green phase durations for the next $W$ time interval. This adaptive Webster's essentially uses the most recent $W$ interval to collect data and assumes the traffic demand will be approximately the same during the next $W$ interval. The selection of $W$ is important and exhibits various trade-offs, smaller values allow for more frequent adaptations to changing traffic demands at the risk of instability while larger values adapt less frequently but allow for increased stability. Pseudo-code for the Webster's traffic signal controller is presented in Algorithm \ref{alg:webster}.

\begin{algorithm}
	\caption{Webster's Algorithm}\label{alg:webster}	
	\begin{algorithmic}[1]
	\Procedure{Webster}{$c_{min}, c_{max}, s, F, R$}
	\State{\#$compute$ $critical$ $lanes$ $for$ $each$ $phase$}
	\State $Y = $ \{ \textrm{max}(\{$\frac{F_l}{s}$ \textbf{for} $l$ \textbf{in} $L_{p,inc}$ \}) \textbf{for} $p$ \textbf{in} $P$ \}
	\State{\#$compute$ $cycle$ $length$}
	\State $C = \frac{(1.5*R)+5}{1.0-\sum Y}$
    \If {$C < c_{min}$}
	\State $C = c_{min}$
	\ElsIf {$C > c_{max}$}
	\State $C = c_{max}$
	\EndIf
	\State $G = C - R$
	\State{\#$allocate$ $green$ $time$ $proportional$ $to$ $flow$}
	\State \textbf{return} $C$, \{ $G\frac{y}{\sum Y}$ \textbf{for} $y$ \textbf{in} $Y$ \}
	\EndProcedure
	\end{algorithmic}
	\end{algorithm}
\par
In Algorithm \ref{alg:webster}, $F$ represents the set of phase flows collected over the most recent $W$ interval and $R$ represents the total cycle lost time. In addition to the time interval hyperparameter $W$, the adaptive Webster's algorithm also has hyperparameters defining a minimum cycle duration $c_{min}$, maximum cycle duration $c_{max}$ and lane saturation flow rate $s$.

\subsubsection{Max-pressure}
The Max-pressure algorithm develops an acyclic, dynamic phase length traffic signal controller. The Max-pressure algorithm models vehicles in lanes as a substance in a pipe and enacts control in a manner which attempts to maximize the relief of pressure between incoming and outgoing lanes \cite{varaiya2013max}. For a given green phase $p$, the pressure is defined in (\ref{eq:pressure}).

	\begin{equation}\label{eq:pressure}
		\textrm{Pressure($p$) } = \sum_{l\in L_{p,inc}} |V_l| - \sum_{l\in L_{p,out}} |V_l| 
	\end{equation}
\par
Where $L_{p,inc}$ represents the set of incoming lanes with green movements in phase $p$ and $L_{p,out}$ represents the set of outgoing lanes from all incoming lanes in $L_{p,inc}$.
\par
Pseudo-code for the Max-pressure traffic signal controller is presented in Algorithm \ref{alg:maxpressure}.
	
	\begin{algorithm}
	\caption{Max-pressure Algorithm}\label{alg:maxpressure}	
	\begin{algorithmic}[1]
	\Procedure{Maxpressure}{$g_{min}, t_{p}, P$}
	\If {$t_{p} < g_{min}$}
	\State $t_{p} = t_{p} + 1$
	\Else
	\State $t_{p} = 0$
	\State{\#$next$ $phase$ $has$ $largest$ $pressure$}
    \State \textbf{return} \textrm{argmax}(\{	\textrm{Pressure}($p$) \textbf{for} $p$ \textbf{in} $P$ \})
	\EndIf
	\EndProcedure
	\end{algorithmic}
	\end{algorithm}
\par
In Algorithm \ref{alg:maxpressure}, $t_p$ represents the time spent in the current phase. The Max-pressure algorithm requires a minimum green time hyperparameter $g_{min}$ which ensures a newly enacted phase has a minimum duration.
\subsubsection{Self Organizing Traffic Lights}
Self-organizing traffic lights (SOTL) \cite{gershenson2004self,cools2013self,goel2017self} develop a cycle-based, dynamic phase length traffic signal controller based on self-organizing principles, where a ``...self-organizing system would be one in which elements are designed to dynamically and autonomously solve a problem or perform a function at the system level.'' \cite[p.~2]{cools2013self}. 

\par
Pseudo-code for the SOTL traffic signal controller is presented in Algorithm \ref{alg:sotl}.
	
	\begin{algorithm}
	\caption{SOTL Algorithm}\label{alg:sotl}	
	\begin{algorithmic}[1]
	\Procedure{SOTL}{$t_{p}, g_{min}, \theta, \omega, \mu$}
	\State{\#$accumulate$ $red$ $phase$ $vehicles$ $time$ $integral$}
	\State{$\kappa = \kappa + \sum_{l\in L_{inc} - L_{p,inc}} |V_l|$}
	\If {$t_{p} > g_{min}$}
	  \State{\#$vehicles$ $approaching$ $in$ $current$ $green$ $phase$}
	  \State{\#$< \omega$ $distance$ $of$ $stop$ $line$}
	  \State $n = \sum_{l\in L_{p,inc}} |V_l|$
	  \State{\#$only$ $consider$ $phase$ $change$ $if$ $no$ $platoon$}
	  \State{\#$or$ $too$ $large$ $n>\mu$}
	  \If {$n > \mu \textbf{ or } n == 0$}
	     \If {$\kappa > \theta$}
	        \State {$\kappa = 0$}
	        \State{\#$next$ $phase$ $in$ $cycle$}
	        \State{$i=i+1$}
	        \State \textbf{return} $P_{i\textrm{mod}|P|}$
	     \EndIf
	  \EndIf
	\EndIf
	\EndProcedure
	\end{algorithmic}
	\end{algorithm}
\par

The SOTL algorithm functions by changing lights according to a vehicle-time integral threshold $\theta$ constrained by a minimum green phase duration $g_{min}$. Additionally, small (i.e. $n < \mu$) vehicle platoons are kept together by preventing a phase change if sufficiently close (i.e., at a distance $< \omega$) to the stop line.

\subsection{Learning Traffic Signal Controllers}

\par
Reinforcement learning uses the framework of Markov Decision Processes to solve goal-oriented, sequential decision-making problems by repeatedly acting in an environment. At discrete points in time $t$, a reinforcement learning agent observes the environment state $s_t$ and then uses a policy $\pi$ to determine an action $a_t$. After implementing its selected action, the agent receives feedback from the environment in the form of a reward $r_t$ and observes a new environment state $s_{t+1}$. The reward quantifies how `well' the agent is achieving its goal (e.g., score in a game, completed tasks). This process is repeated until a terminal state $s_{terminal}$ is reached, and then begins anew. The return $ G_t=\sum_{k=0}^{k=T} \gamma^k r_{t+k}$ is the accumulation of rewards by the agent over some time horizon $T$, discounted by $\gamma \in [0,1)$. The agent seeks to maximize the expected return $\mathbb{E}[G_t]$ from each state $s_t$. The agent develops an optimal policy $\pi^*$ to maximize the return. 
\par
There are many techniques for an agent to learn the optimal policy, however, most of them rely on estimating value functions. Value functions are useful to estimate future rewards. State value functions $V^\pi(s) = \mathbb{E}[G_t|s_t=s]$ represent the expected return starting from state $s$ and following policy $\pi$. Action value functions $Q^\pi(s,a)= \mathbb{E}[G_t|s_t=s,a_t=a]$ represent the expected return starting from state $s$, taking action $a$ and following policy $\pi$. In practice, value functions are unknown and must be estimated using sampling and function approximation techniques. Parametric function approximation, such as neural networks, use a set of parameters $\theta$ to estimate an unknown function $f(x|\theta) \approx f(x)$. To develop accurate approximations, the function parameters must be developed with some optimization technique.
\par
Experiences are tuples $e_t=(s_t,a_t,r_t,s_{t+1})$ that represent an interaction between the agent and the environment at time $t$. A reinforcement learning agent interacts with its environment in trajectories or sequences of experiences $e_{t}, e_{t+1}, e_{t+2},...$. Trajectories begin in an initial state $s_{init}$ and end in a terminal state $s_{terminal}$. To accurately estimate value functions, experiences are used to optimize the parameters. If neural network function approximation is used, the parameters are optimized using experiences to perform gradient-based techniques and backpropagation  \cite{linnainmaa1976taylor,rumelhart1986learning}. Additional technical details regarding the proposed reinforcement learning adaptive traffic signal controllers can be found in the Appendix.

\begin{figure*}[!t]
\centering
\includegraphics[scale=0.475]{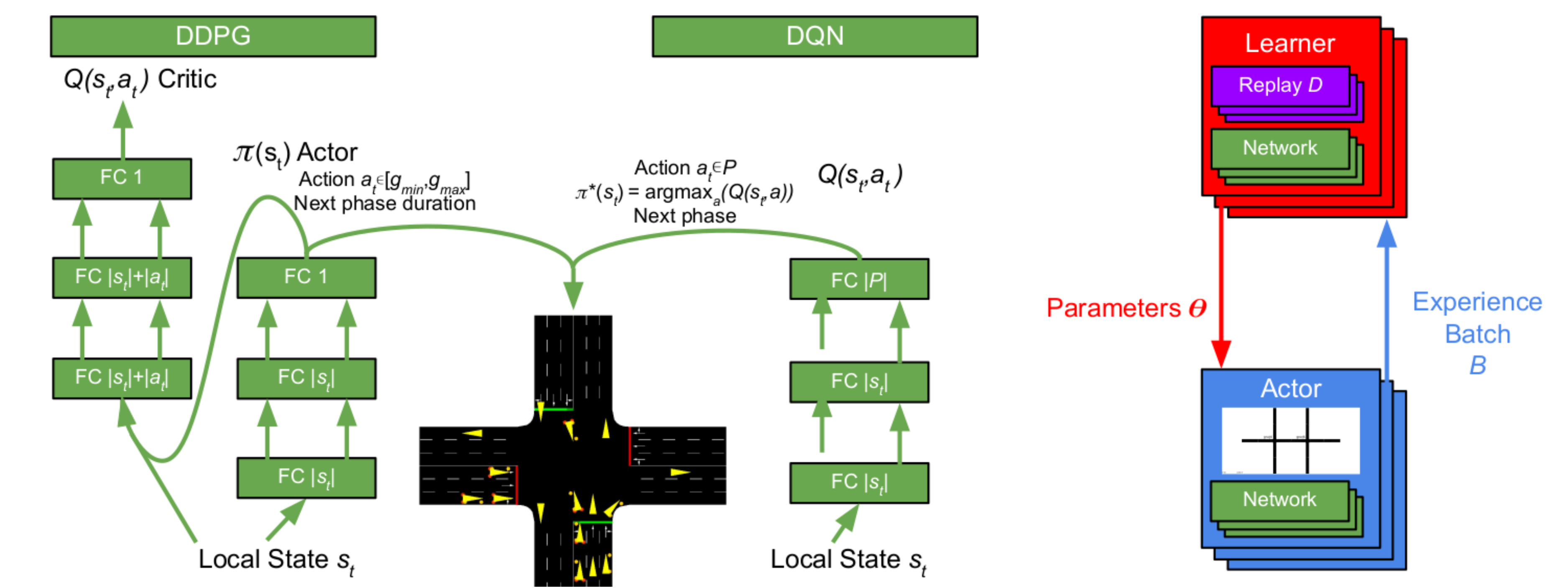}
\caption{ Adaptive traffic signal control DDPG and DQN neural network agents (left) and distributed acting, centralized learning architecture (right) composed of actors and learners. Each actor has one SUMO network as an environment and neural networks for all intersections. Each learner is  assigned a subset of intersections at the beginning of training and is only responsible for computing parameter updates for their assigned intersections, effectively distributing the computation load for learning. However, learners distribute parameter updates to all actors.}
\label{fig:model_arch}
\end{figure*}

\par
	To train reinforcement learning controllers for all intersections, a distributed acting, centralized learning architecture is developed \cite{mnih2016asynchronous,horgan2018distributed,espeholt2018impala}. Using parallel computing, multiple actors and learners are created, illustrated in Figure \ref{fig:model_arch}. Actors have their own instance of the traffic simulation and neural networks for all intersections. Learners are assigned a subset of all intersections, for each they have a neural network and an experience replay buffer $D$. Actors generate experiences $e_t$ for all intersections and send them to the appropriate learner. Learners only receive experiences for their assigned subset of intersections. The learner stores the experiences in an experience replay buffer, which is uniformly sampled for batches to optimize the neural network parameters. After computing parameter updates, learners send new parameters to all actors.
	\par
	There are many benefits to this architecture, foremost being that it makes the problem feasible; because there are hundreds of agents, distributing computation across many actors and learners is necessary to decrease training time. Another benefit is experience diversity, granted by multiple environments and varied exploration rates.

\subsection{DQN}
\par
    The proposed DQN traffic signal controller enacts control by choosing the next green phase without utilizing a phase cycle. This acyclic architecture is motivated by the observation that enacting phases in a repeating sequence may be contributing to sub-optimal control policy. After the DQN has selected the next phase, it is enacted for a fixed duration known as an action repeat $a_{repeat}$. After the phase has been enacted for the action repeat duration, a new phase is selected acyclically. 
    \subsubsection{State}
     The proposed state observation for the DQN is a combination of the most recent green phase and the density and queue of incoming lanes at the intersection at time $t$. Assume each intersection has a set $L$ of incoming lanes and a set $P$ of green phases. The state space is then defined as $S\in (\mathbb{R}^{2|L|}\times\mathbb{B}^{|P|+1})$. The density and queue of each lane are normalized to the range $[0,1]$ by dividing by the lane's jam density $k_j$. The most recent phase is encoded as a one-hot vector $\mathbb{B}^{|P|+1}$, where the plus one encodes the all-red clearance phase.
    \subsubsection{Action}
    The proposed action space for the DQN traffic signal controller is the next green phase. The DQN selects one action from a discrete set, in this model one of the many possible green phases $a_t\in P$. After a green phase has been selected, it is enacted for a duration equal to the action repeat $a_{repeat}$.
    \subsubsection{Reward}
	The reward used to train the DQN traffic signal controller is a function of vehicle delay. Delay $d$ is the difference in time between a vehicle's free-flow travel time and actual travel time. Specifically, the reward is the negative sum of all vehicles' delay at the intersection, defined in (\ref{eq:reward}):
	
	\begin{equation}\label{eq:reward}
		r_t = -\sum_{v\in V} d^v_t
	\end{equation}	
	\par
	Where $V$ is the set of all vehicles on incoming lanes at the intersection, and $d^t_v$ is the delay of vehicle $v$ at time $t$. Defined in this way, the reward is a punishment, with the agent's goal to minimize the amount of punishment it receives. Each intersection saves the reward with the largest magnitude experienced to perform minimum reward normalization $\frac{r_t}{|r_{min}|}$ to scale the reward to the range $[-1, 0]$ for stability.

    \subsubsection{Agent Architecture}
     The agent approximates the action-value $Q$ function with a deep artificial neural network. The action-value function $Q$ is two hidden layers of $3(|s_t|)$ fully connected neurons with exponential linear unit (ELU) activation functions and the output layer is $|P|$ neurons with linear activation functions. The $Q$ function's input is the local intersection state $s_t$. A visualization of the DQN is presented in Fig. \ref{fig:model_arch}. 

\subsection{DDPG Traffic Signal Controller}
    \par
    The proposed DDPG traffic signal controller implements a cycle with dynamic phase durations. This architecture is motivated by the observation that cycle-based policies can maintain fairness and ensure a minimum quality of service between all intersection users. Once the next green phase has been determined using the cycle, the policy $\pi$ is used to select its duration. Explicitly, the reinforcement learning agent is learning how long in duration to make the next green phase in the cycle to maximize its return. Additionally, the cycle skips phases when no vehicles are present on incoming lanes.
    \subsubsection{Actor State}
    \par
    The proposed state observation for the actor is a combination of the current phase and the density and queue of incoming lanes at the intersection at time $t$. The state space is then defined as $S\in (\mathbb{R}^{2|L|}\times\mathbb{B}^{|P|+1})$. The density and queue of each lane are normalized to the range $[0,1]$ by dividing by the lane's jam density $k_j$. The current phase is encoded as a one-hot vector $\mathbb{B}^{|P|+1}$, where the plus one encodes the all-red clearance phase.
    \subsubsection{Critic State}
    \par
    The proposed state observation for the critic combines the state $s_t$ and the actor's action $a_t$, depicted in Figure \ref{fig:model_arch}.
    \subsubsection{Action}
    \par
    The proposed action space for the adaptive traffic signal controller is the duration of the next green phase in seconds. The action controls the duration of the next phase; there is no agency over what the next phase is, only on how long it will last. The DDPG algorithm produces a continuous output, a real number over some range $a_t\in \mathbb{R}$. Since the DDPG algorithm outputs a real number and the phase duration is defined in intervals of seconds, the output is rounded to the nearest integer. In practice, phase durations are bounded by minimum time $g_{min}$ and a maximum time $g_{max}$ hyperparameters to ensure a minimum quality of service for all users. Therefore the agent selects an action $ \{a_t \in \mathbb{Z} | g_{min} \leq a_t \leq g_{max} \}$ as the next phase duration.
    \subsubsection{Reward}
    The reward used to train the DDPG traffic signal controller is the same delay reward used by the DQN traffic signal controller defined in (\ref{eq:reward}).
    \subsubsection{Agent Architecture}
    The agent approximates the policy $\pi$ and action-value $Q$ function with deep artificial neural networks. The policy function is two hidden layers of $3|s_t|$ fully connected neurons, each with batch normalization and ELU activation functions, and the output layer is one neuron with a hyperbolic tangent activation function. The action-value function $Q$ is two hidden layers of $3(|s_t|+|a_t|)$ fully connected neurons with batch normalization and ELU activation functions and the output layer is one neuron with a linear activation function. The policy's input is the intersection's local traffic state $s_t$ and the action-value function's input is the local state concatenated with the local action $s_t+a_t$. The action-value $Q$ function also uses a $L_2$ weight regularization of $\lambda=0.01$. 
    
    \par
	By deep reinforcement learning standards the networks used are not that deep, however, their architecture is selected for simplicity and they can easily be modified within the framework. Simple deep neural networks were also implemented to allow for future scalability, as the proposed framework can be deployed to any SUMO network - to reduce the computational load the default networks are simple.

\begin{figure}[!t]
\centering
 \includegraphics[scale=0.45]{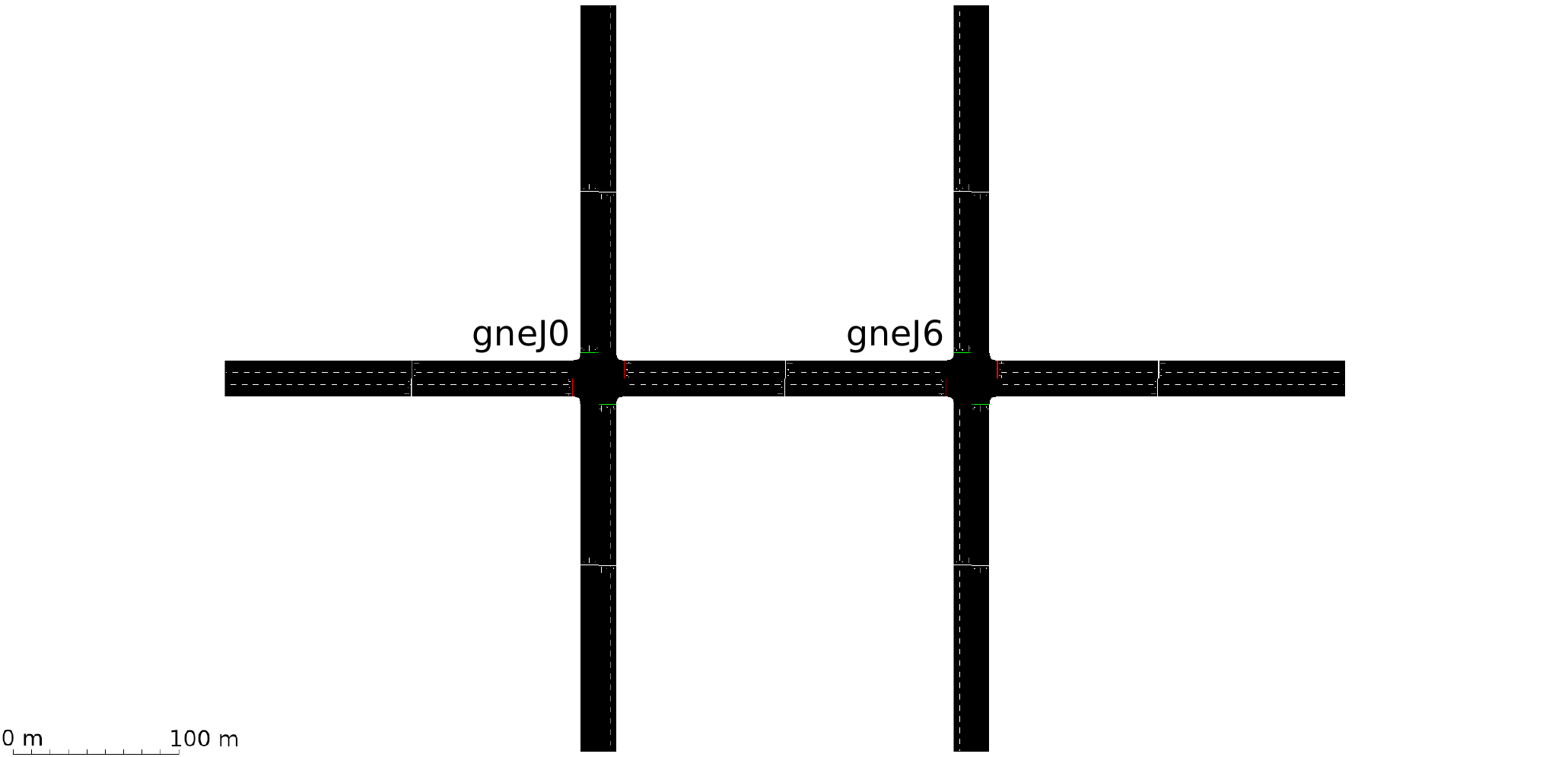}
    \caption{Two intersection SUMO network used for hyperparameter experiments. In addition to this two intersection network, a single, isolated intersection is also included with the framework.}   
    \label{fig:double_network}
\end{figure}

\section{Experiments}
\subsection{Hyperparameter Optimization}
To demonstrate the capabilities of the proposed framework, experiments are conducted on optimizing adaptive traffic signal control hyperparameters. The framework is for use with the SUMO traffic microsimulator \cite{SUMO2012}, which was used to evaluate the developed adaptive traffic signal controllers. Understanding how sensitive any specific adaptive traffic signal controller's performance is to changes in hyperparameters is important to instill confidence that the solution is robust. Determining optimal hyperparameters is necessary to ensure a balanced comparison between adaptive traffic signal control methods.
\par
\begin{figure*}[!t]
\centering
\includegraphics[scale=0.25]{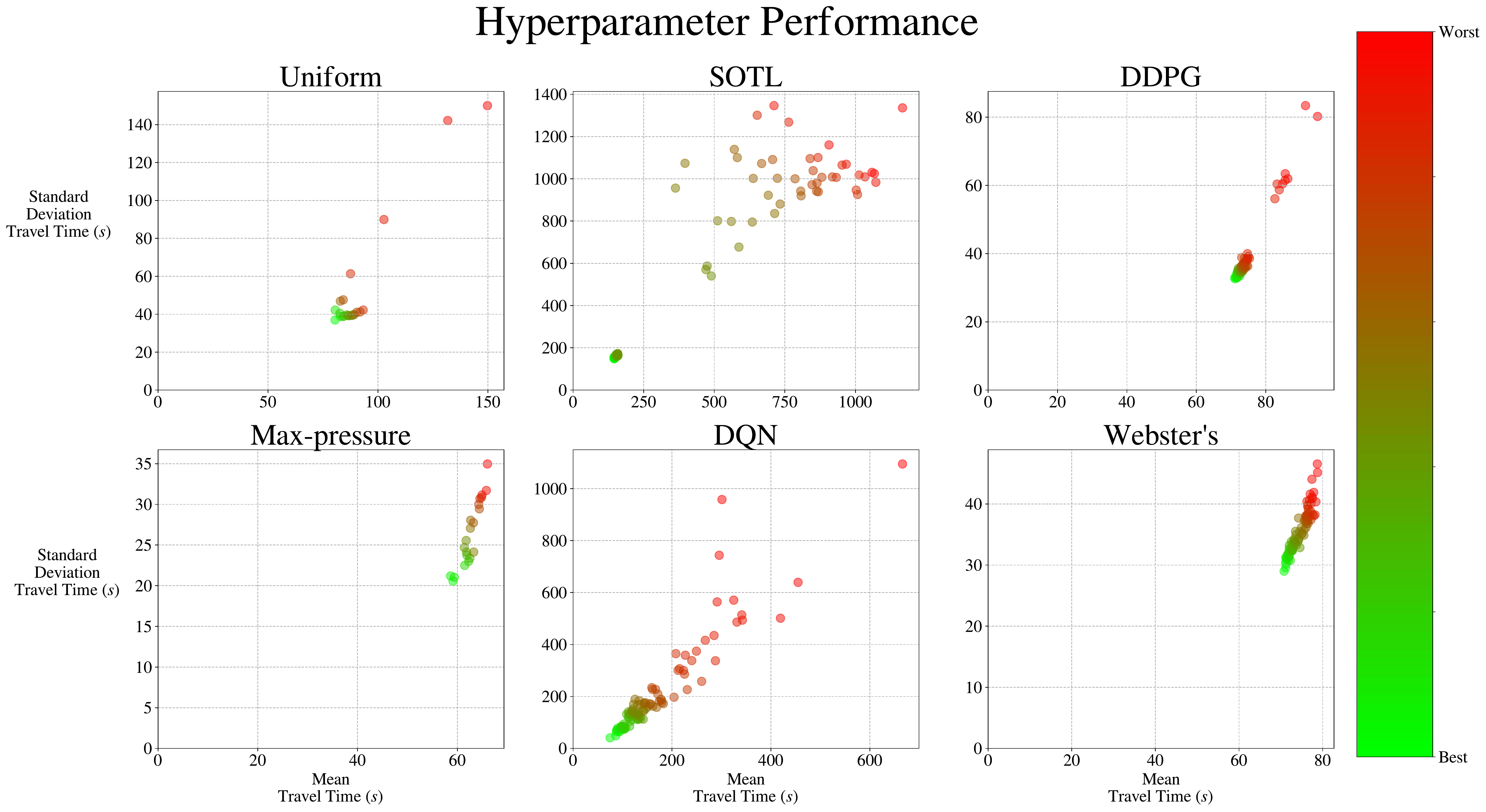}
\caption{Individual hyperparameter results for each traffic signal controller. Travel time is used as a measure of effectiveness and is estimated for each hyperparameter from eight simulations with random seeds in units of seconds ($s$). The coloured dots gradient from green (best) to red (worst) orders the hyperparameters by the sum of the travel time mean and standard deviation. Note differing scales between graph axes making direct visual comparison biased.}
\label{fig:tsc_hp}
\end{figure*}
\begin{figure*}[!t]
\centering
 \includegraphics[scale=0.27]{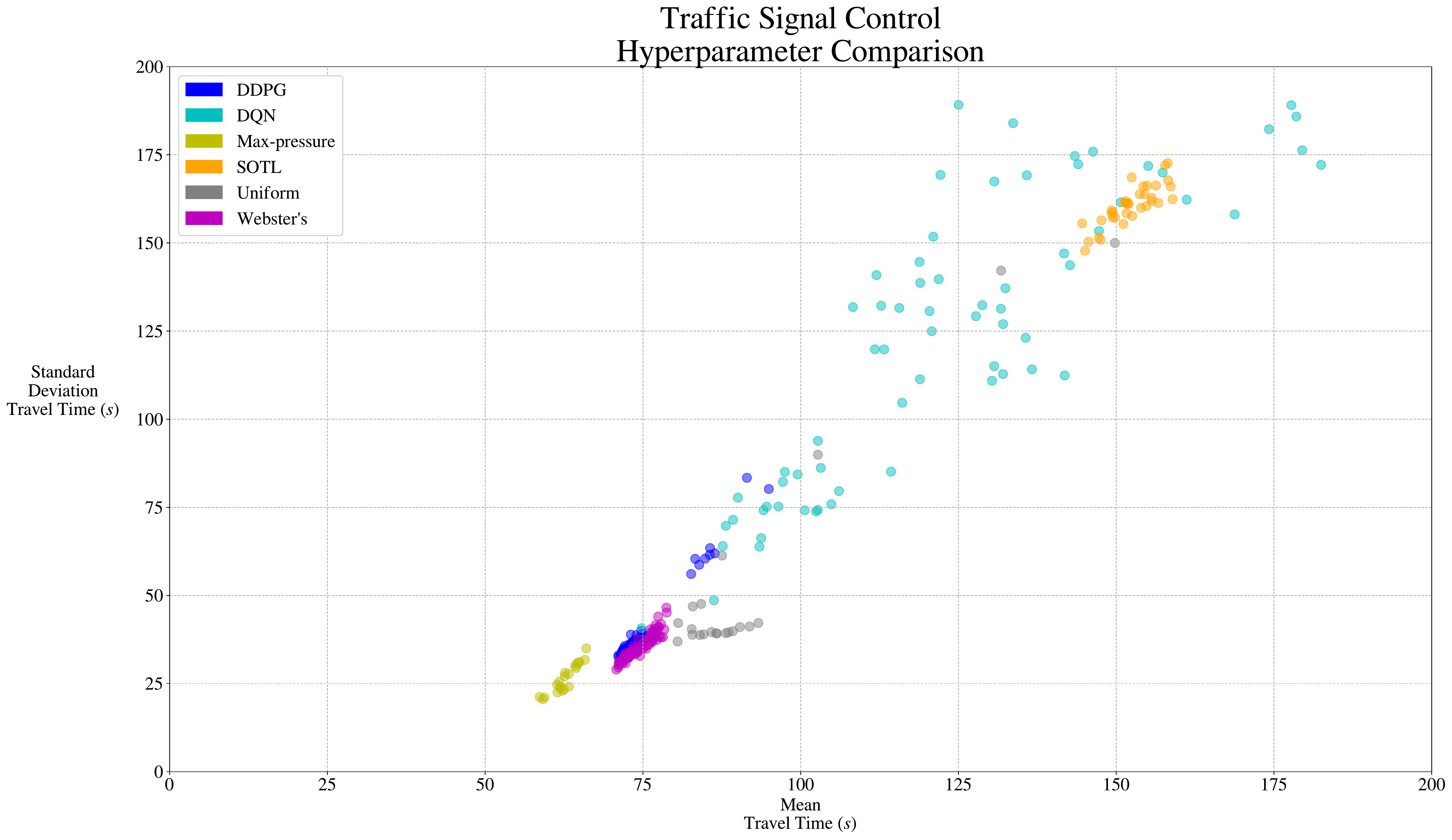}
    \caption{Comparison of all traffic signal controller hyperparameter travel time performance. Note both vertical and horizontal axis limits have been clipped at 200 to improve readability.}   
    \label{fig:hp}
\end{figure*}

Using the hyperparameter optimization script included in the framework, a grid search is performed with the implemented controllers' hyperparameters on a two intersection network, shown in Fig. \ref{fig:double_network}  under a simulated three hour dynamic traffic demand scenario. The results for each traffic signal controller are displayed in Fig. \ref{fig:tsc_hp} and collectively in Fig. \ref{fig:hp}.
\par
As can be observed in Fig. \ref{fig:tsc_hp} and Fig. \ref{fig:hp} the choice of hyperparameter significantly impacts the performance of the given traffic signal controller. As a general trend observed in Fig. \ref{fig:tsc_hp}, methods with larger numbers of hyperparameters (e.g., SOTL, DDPG, DQN) exhibit greater performance variance than methods with fewer hyperparameters (e.g., Max-pressure). Directly comparing methods in Fig. \ref{fig:hp} demonstrates non-learning adaptive traffic signal control methods (e.g., Max-pressure, Webster's) robustness to hyperparameter values and high performance (i.e., lowest travel time). Learning-based methods exhibit higher variance with changes in hyperparameters, DQN more so than DDPG. In the following section, the best hyperparameters for each adaptive traffic signal controller will be used to further investigate and compare performance.

\subsection{Optimized Adaptive Traffic Signal Controllers}

\begin{figure*}[!t]
\centering
\includegraphics[scale=0.27]{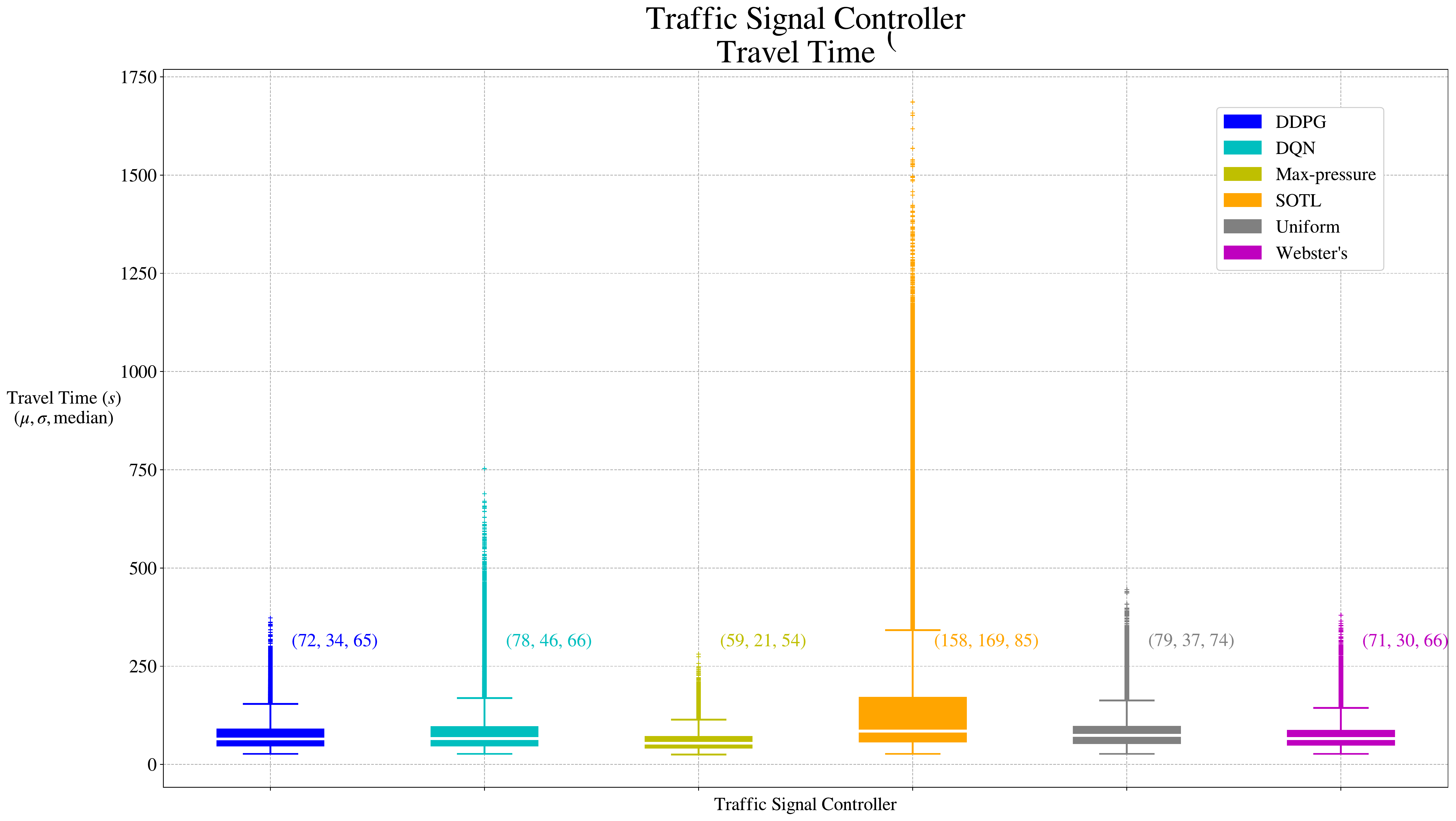}
\caption{Boxplots depicting the distribution of travel times for each traffic signal controller. The solid white line represents the median, solid coloured box the interquartile range (IQR) (i.e., from first (Q1) to third quartile (Q3)), solid coloured lines the Q1-$1.5$IQR and Q3+$1.5$IQR and coloured crosses the outliers.}
\label{fig:travel_time}
\end{figure*}

\par
\begin{figure*}[!t]
\centering
 \includegraphics[scale=0.27]{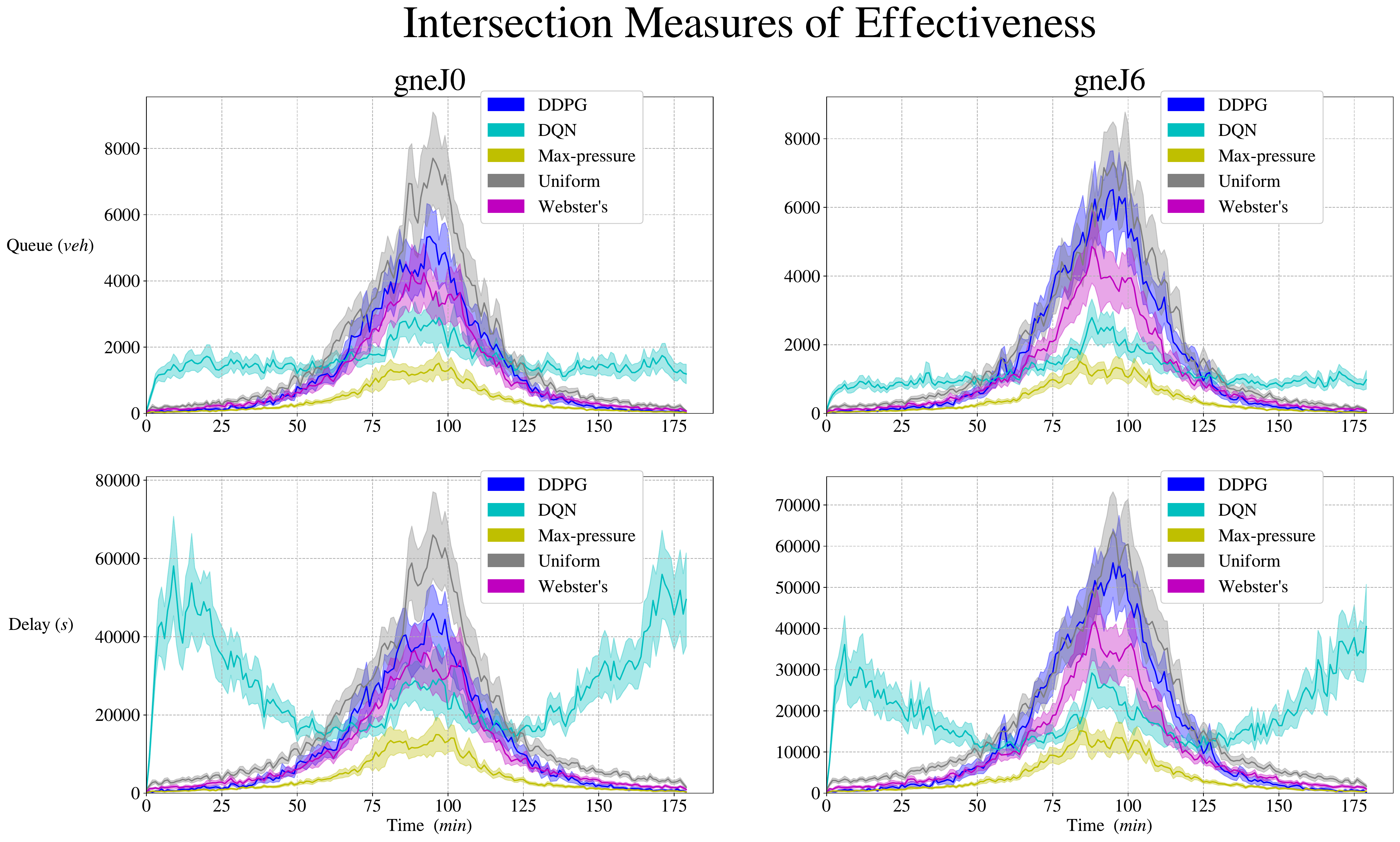}
    \caption{Comparison of traffic signal controller individual intersection queue and delay measure of effectiveness in units of vehicles $(veh)$ and seconds ($s$). Solid coloured lines represent the mean and shaded areas represent the $95\%$ confidence interval. SOTL has been omitted to improve readability since its Queue and Delay values are exclusively outside the graph range.}   
    \label{fig:intersection_moe}
\end{figure*}

Using the optimized hyperparameters all traffic signal controllers are subjected to an additional $32$ simulations with random seeds to estimate their performance, quantified using network travel time, individual intersection queue and delay measures of effectiveness (MoE). Results are presented in Fig. \ref{fig:travel_time} and Fig. \ref{fig:intersection_moe}.
\par
Observing the travel time boxplots in Fig. \ref{fig:travel_time}, the SOTL controller produces the worst results, exhibiting a mean travel time almost twice the next closest method and with many significant outliers. The Max-pressure algorithm achieves the best performance, with the lowest mean and median along with the lowest standard deviation. The DQN, DDPG, Uniform and Webster's controllers achieve approximately equal performance, however, the DQN controller has significant outliers, indicating some vehicles experience much longer travel times than most.
\par
Each intersection's queue and delay MoE with respect to each adaptive traffic signal controller is presented in Fig. \ref{fig:intersection_moe}. The results are consistent with previous observations from the hyperparameter search and travel time data, however, the reader's attention is directed to comparing the performance of DQN and DDPG in Fig. \ref{fig:intersection_moe}. The DQN controller performs poorly (i.e., high queues and delay) at the beginning and end of the simulation when traffic demand is low. However, at the demand peak, the DQN controller performs just as well, if not a little better, than every method except the Max-pressure controller. Simultaneously considering the DDPG controller, the performance is opposite the DQN controller. The DDPG controller achieves relatively low queues and delay at the beginning and end of the simulation and then is bested by the DQN controller in the middle of the simulation when the demand peaks. This performance difference can potentially be understood by considering the difference between the DQN and DDPG controllers. The DQN's ability to select the next phase acyclically under high traffic demand may allow it to reduce queues and delay more than the cycle constrained DDPG controller. However, it is curious that under low demands the DQN controller performance suffers when it should be relatively simple to develop the optimal policy. The DQN controller may be overfitting to the periods in the environment when the magnitude of the rewards are large (i.e., in the middle of the simulation when the demand peaks) and converging to a policy that doesn't generalize well to the environment when the traffic demand is low. The author's present these findings to reader's and suggest future research investigate this and other issues to understand the performance difference between reinforcement learning traffic signal controllers. Understanding the advantages and disadvantages of a variety of controllers can provide insight into developing future improvements.

\section{Conclusion \& Future Work}
Learning and non-learning adaptive traffic signal controllers have been developed within an optimized framework for the traffic microsimulator SUMO for use by the research community. The proposed framework's capabilities were demonstrated by studying adaptive traffic signal control algorithm's sensitivity to hyperparameters, which was found to be sensitive with hyperparameter rich controllers (i.e., learning) and relatively insensitive with hyperparameter sparse controllers (i.e., heuristics). Poor hyperparameters can drastically alter the performance of an adaptive traffic signal controller, leading researchers to erroneous conclusions about an adaptive traffic signal controller's performance. This research provides evidence that dozens or hundreds of hyperparameter configurations may have to be tested before selecting the optimal one.
\par
Using the optimized hyperparamters, each adaptive controller's performance was estimated and the Max-pressure controller was found to achieve the best performance, yielding the lowest travel times, queues and delay. This manuscript's research provides evidence that heuristics can offer powerful solutions even compared to complex deep-learning methods. This is not to suggest that this is definitively the case in all environments and circumstances. The authors' hypothesize that learning-based controllers can be further developed to offer improved performance that may yet best the non-learning, heuristic-based methods detailed in this research. Promising extensions that have improved reinforcement learning in other applications and may do the same for adaptive traffic signal control include richer function approximators \cite{bellemare2017distributional,dabney2018distributional,dabney2018implicit} and reinforcement learning algorithms \cite{sutton1999between,bacon2017option,sharma2017learning}.
\par
The authors intend for the framework to grow, with the addition of more adaptive traffic signal controllers and features. In its current state, the framework can already aid adaptive traffic signal control researchers rapidly experiment on a SUMO network of their choice. Acknowledging the importance of optimizing our transportation systems, the authors hope this research helps others solve practical problems.

\appendices
\section{DQN}

Deep Q-Networks \cite{mnih2015human} combine Q-learning and deep neural networks to produce autonomous agents capable of solving complex tasks in high-dimensional environments. Q-learning \cite{watkins1992q} is a model-free, off-policy, value-based temporal difference \cite{sutton1988learning} reinforcement learning algorithm which can be used to develop an optimal discrete action space policy for a given problem. Like other temporal difference algorithms, Q-learning uses bootstrapping \cite{sutton1998reinforcement} (i.e., using an estimate to improve future estimates) to develop an action-value function $Q(s,a)$ which can estimate the expected return of taking action $a$ in state $s$ and acting optimally thereafter. If the $Q$ function can be estimated accurately it can be used to derive the optimal policy $\pi^* = \textrm{argmax}_aQ(s,a)$. In DQN the $Q$ function is approximated with a deep neural network. The DQN algorithm utilizes two techniques to ensure stable development of the $Q$ function with DNN function approximation - a target network and experience replay. Two parameter sets are used when training a DQN, online $\theta$ and target $\theta^{\prime}$. The target parameters $\theta^{\prime}$ are used to stabilize the return estimates when performing updates to the neural network and are periodically changed to the online parameters $\theta^{\prime} = \theta$ at a fixed interval. The experience replay is a buffer which stores the most recent $D$ experience tuples to create a slowly changing dataset. The experience replay is uniformly sampled from for experience batches to update the $Q$ function online parameters $\theta$.
	
	\par 
	Training a deep neural network requires a loss function, which is used to determine how to change the parameters to achieve better approximations of the training data. Reinforcement learning develops value functions (e.g., neural networks) using experiences from the environment.
	
	\par
	The DQN's loss function is the gradient of the mean squared error of the return $G_t$ target $y_t$ and the prediction, defined in (\ref{eq:dqnloss}). 
	\begin{equation}\label{eq:dqnloss}
	\begin{aligned}
		y_t &= r_t+\gamma Q(s_{t+1}, \textrm{argmax}_a Q(s_{t+1}, a|\theta^\prime)|\theta^\prime) \\
		L_{\textrm{DQN}}(\theta) &= (y_t - Q(s_t,a_t|\theta))^2
	\end{aligned}
	\end{equation}

\section{DDPG}

	Deep deterministic policy gradients \cite{lillicrap2015continuous} are an extension of DQN to continuous action spaces. Similiar to Q-learning, DDPG is a model-free, off-policy reinforcement learning algorithm. The DDPG algorithm is an example of actor-critic learning, as it develops a policy function $\pi(s|\phi)$ (i.e., actor) using an action-value function $Q(s,a|\theta)$ (i.e., critic). The actor interacts with the environment and modifies its behaviour based on feedback from the critic. 
	
	\par
	The DDPG critic's loss function is the gradient of the mean squared error of the return $G_t$ target $y_t$ and the prediction, defined in (\ref{eq:criticloss}). 
	\begin{equation}\label{eq:criticloss}
	\begin{aligned}
		y_t &= r_t+\gamma Q(s_{t+1}, \pi(s_{t+1}|\phi^\prime) |\theta^\prime) \\
		L_{\textrm{Critic}}(\theta) &= (y_t - Q(s_t,a_t|\theta))^2 
	\end{aligned}
	\end{equation}
	 
	\par	
	The DDPG actor's loss function is the sampled policy gradient, defined in (\ref{eq:actorloss}).
    \begin{equation}\label{eq:actorloss}
	\begin{aligned}
	L_{\textrm{Actor}}(\theta) &= \nabla_\theta Q(s_t, \pi(s_t|\phi)|\theta) 
	\end{aligned}
	\end{equation}
    \par
    Like DQN, DDPG uses two sets of parameters, online $\theta$ and target $\theta^\prime$, and experience replay \cite{lin1992self} to reduce instability during training. DDPG performs updates on the parameters for both the actor and critic by uniformly sampling batches of experiences from the replay. The target parameters are slowly updated towards the online parameters according to $\theta^\prime = (1-\tau)\theta^\prime + (\tau)\theta$ after every batch update.

\section{Technical}
\par
Software used include SUMO 1.2.0 \cite{SUMO2012}, Tensorflow 1.13 \cite{tensorflow2015-whitepaper}, SciPy \cite{scipy}  and public code \cite{ddpg2019}. The neural network parameters were initialized with He \cite{he2015delving} and optimized using Adam \cite{kingma2014adam}.
\par 
To ensure intersection safety, two second yellow change and three second all-red clearance phases were inserted between all green phase transitions. For the DQN and DDPG traffic signal controllers, if no vehicles are present at the intersection, the phase defaults to all-red, which is considered a terminal state $s_{terminal}$. Each intersection's state observation is bounded by $150$ m (i.e., the queue and density are calculated from vehicles up to a maximum of $150$ m from the intersection stop line).
\section*{Acknowledgments}
\par
This research was enabled in part by support in the form of computing resources provided by SHARCNET (\url{www.sharcnet.ca}), their McMaster University staff and Compute Canada (\url{www.computecanada.ca}). 

\ifCLASSOPTIONcaptionsoff
  \newpage
\fi



%
\bibliographystyle{IEEEtran}
\bibliography{references}
%

\begin{IEEEbiography}[{\includegraphics[height=1.2in,keepaspectratio]{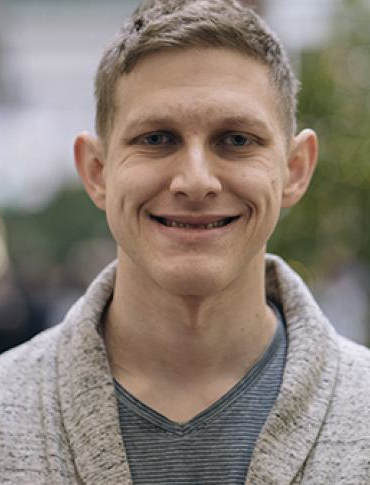}}]{Wade Genders}
earned a Software B.Eng. \& Society in 2013, Civil M.A.Sc. in 2014 and Civil Ph.D in 2018 from McMaster University. His research interests include traffic signal control, intelligent transportation systems, machine learning and artificial intelligence.
\end{IEEEbiography}

\begin{IEEEbiography}[{\includegraphics[height=1.2in,keepaspectratio]{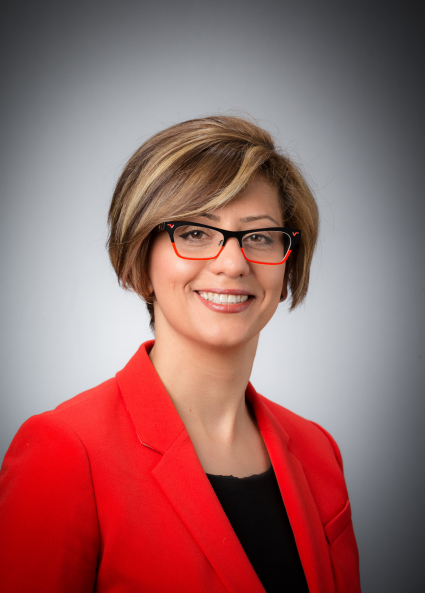}}]{Saiedeh Razavi}
Saiedeh Razavi is the inaugural Chair in Heavy Construction, Director of the McMaster Institute for Transportation and Logistics and Associate Professor at the Department of Civil Engineering at McMaster University. Dr. Razavi has a multidisciplinary background and considerable experience in collaborating and leading national and international multidisciplinary team-based projects in sensing and data acquisition, sensor technologies, data analytics, data fusion and their applications in safety, productivity, and mobility of transportation, construction,
and other systems. She combines several years of industrial experience with academic teaching and research. Her formal education includes degrees in Computer Engineering (B.Sc), Artificial Intelligence (M.Sc) and Civil Engineering (Ph.D.). Her research, funded by Canadian council (NSERC), as well as the ministry of Transportation of Ontario, focuses on connected and automated vehicles, on smart and connected work zones and on computational models for improving safety and productivity of highway construction. Dr. Razavi brings together the private and public sectors with academia for the development of high quality research in smarter mobility, construction and logistics. She has received several awards including McMaster’s Student Union Merit Award for Teaching, the Faculty of Engineering Team Excellent Award, and the Construction Industry Institute best poster award.
\end{IEEEbiography}

\end{document}